# Unusual Kondo Behavior in the Indium-Rich Heavy Fermion Antiferromagnet $Ce_3Pt_4In_{13}$

M.F. Hundley, J.L. Sarrao, J.D. Thompson, R. Movshovich, and M. Jaime

*Los Alamos National Laboratory, Los Alamos, NM 87545*

C. Petrovic* and Z. Fisk

*National High Magnetic Field Laboratory,*

*Florida State University, Tallahassee, FL 32310*

We report the thermodynamic, magnetic, and electronic transport properties of the new ternary intermetallic system $(Ce,La)_3Pt_4In_{13}$. $Ce_3Pt_4In_{13}$ orders antiferromagnetically at 0.95 K while the non-magnetic compound $La_3Pt_4In_{13}$ is a conventional 3.3 K superconductor. Kondo lattice effects appear to limit the entropy associated with the Néel transition to $(1/4)Rln2$ as an electronic contribution to the specific heat of $\gamma \sim 1$ J/mole-Ce $K^2$ is observed at $T_N$; roughly 35% of this $\gamma$ survives the ordering transition. Hall effect, thermoelectric power, and ambient-pressure resistivity measurements confirm this interpretation. These results suggest that RKKY and Kondo interactions are closely balanced in this compound ($T_N \approx T_K$). Contrary to expectations based on the Doniach Kondo necklace model, applied hydrostatic pressure modestly enhances the magnetic ordering temperature with $dT_N/dP = +23$ mK/kbar. As such $Ce_3Pt_4In_{13}$ provides a counterexample to Kondo systems with similar Kondo and RKKY energy scales wherein applied pressure enhances $T_K$ at the expense of the ordered magnetic state.

**I.    Introduction**

The low-temperature magnetic and electronic properties of certain rare-earth and actinide compounds are determined by the competition between RKKY and Kondo interactions.[1,2] While these 4f and 5f compounds usually display localized full-moment paramagnetic behavior near room-temperature, the low-temperature state is usually quite different, showing signs of superconductivity, local-moment magnetic order, or in some cases a paramagnetic groundstate and a highly renormalized electronic mass.[1] Doniach's treatment [3] of a one-dimensional Kondo lattice demonstrated that the ground-state magnetic behavior depends upon the relative strengths of RKKY and Kondo interactions. The strengths of these competing interactions are determined by the product $|J_{ex}N(E_F)|$, where $J_{ex}$ is the magnetic exchange energy and $N(E_F)$ is the density of states at the Fermi energy $E_F$. The phase diagram that results from this simple model describes qualitatively the ground state properties of several correlated electron 4f and 5f compounds.[4,5] For low values of $|J_{ex}N(E_F)|$ the RKKY interaction dominates and an ordered magnetic groundstate with very little discernable carrier mass renormalization results.[6] At the other extreme, where $|J_{ex}N(E_F)|$ is large, Kondo interactions predominate so that no magnetic order occurs and a highly renormalized electronic state is realized at low temperatures.[1] The competition is most evident in compounds with an intermediate value for $|J_{ex}N(E_F)|$; in this case $T_N \sim T_K$ so that both magnetic order and an enhanced electron mass can occur, as is the case with $CePt_2Sn_2$,[7] $CeRhIn_5$,[8] and possibly $CePtSb$.[9] The near-equivalent Kondo and magnetic ordering energy scales in this last regime provides an opportunity to test our understanding of the interplay between these competing interactions.

In this paper we report the physical properties of the previously unreported ternary intermetallic compound $Ce_3Pt_4In_{13}$ that exhibits near-identical Kondo and magnetic energy scales. This compound forms in the same structure as a series of stannide-phase superconductors that were first reported by Remeika.[10] Specific heat, magnetic susceptibility, and electronic transport data indicate that $Ce_3Pt_4In_{13}$ orders antiferromagnetically at $T_N = 0.95$ K. The entropy liberated at $T_N$ is only $(1/4)Rln2$ while the remaining $(3/4)Rln2$ entropy associated with a doublet ground state is spread between 1 and 10 K, suggesting that the ordered doublet is strongly Kondo-compensated. The specific-heat Sommerfeld coefficient $\gamma$, which reflects mass enhancement due to Kondo-like effects, is estimated to be $\gamma \approx 1$ J/mole-Ce $K^2$, corresponding to a Kondo temperature $T_K \approx 4.5$ K. The transport properties below 50 K are dominated by a crystalline electric field (CEF) anomaly centered near 15 K that results from a doublet magnetic level situated 25 K above the ground state level. Following Doniach's treatment of the Kondo necklace model for compounds with nearly identical magnetic and Kondo energy scales, one would assume that applied pressure should enhance the Kondo effect (increase $T_K$) at the expense of the magnetic order (decrease $T_N$). While the CEF anomaly precludes tracking the effects of pressure on $T_K$, pressure actually increases $T_N$, signifying that $Ce_3Pt_4In_{13}$ provides a counterexample to the standard expectation in materials where there is a close balance between $T_K$ and $T_N$.

## II. Experimental Details

Single crystals of $Ce_3Pt_4In_{13}$ and $La_3Pt_4In_{13}$ with typical dimensions of 5x5x1 $mm^3$ were grown from an indium-rich flux. X-ray diffraction on powdered crystals indicates that the samples are single-phased and form in the cubic $Pr_3Rh_4Sn_{13}$ structure (Pm3n) first reported by Rameika, *et al.*[10] This structure has a single rare-earth site with cuboctahedral ($\bar{4}2m$) point symmetry and a Ce-Ce distance of $a_o/2$.[11,12] The lattice parameters for the Ce and La compounds are $a_o = 9.771$ Å and $a_o = 9.812$ Å, respectively.

In order to ensure reproducibility in the electronic transport and specific heat measurements performed on the flux-grown $(Ce,La)_3Pt_4In_{13}$ materials examined in this study, it was imperative that all samples used were free of excess surface indium. To that end, all specimens were prescreened by ensuring that no resistive signature was evident at the $T_c$ of indium (3.4 K). dc magnetic susceptibility measurements were made with a Quantum-Design SQUID magnetometer in a field of 1 kOe over the temperature range 2 K to 350 K. The specific-heat was measured with a small-mass relaxation-time calorimeter for 1.5 K < T < 20 K, and with an adiabatic calorimeter for 0.05 K < T < 1.5 K. Resistivity measurements were made with a conventional four-probe low-frequency ac technique, while Hall measurements were done in a 10 kOe field with a conventional four-probe Hall geometry and a low-frequency ac bridge. The thermoelectric power (TEP) was measured with a slowly varying gradient technique. In all cases, electrical sample contacts were made with silver conductive paint or silver epoxy. Hydrostatic pressure-dependent resistivity measurements were performed in a clamped BeCu pressure cell [13] with Flourinert-75 as the pressure medium; the operating pressures were determined from a Sn manometer.

### III. Results

The temperature-dependent magnetic susceptibility $\chi(T)$ is depicted in Fig. 1. The susceptibility exhibits Curie-Weiss behavior $\chi = C/(T-\theta)$ above 200 K with an effective moment of 2.64 $\mu_B$/Ce, which is only slightly larger than the full Hund's rule value of 2.54 $\mu_B$/Ce expected for $Ce^{3+}$. The paramagnetic $\theta$ is -36 K, indicative of antiferromagnetic interactions above 200K. $\chi(T)$ deviates from the Curie-Weiss law below 200 K in a manner which suggests that the J = 5/2 $Ce^{3+}$ manifold is split by crystal field effects. Below 10 K (see the inset to Fig. 1) $\chi$ saturates in a manner characteristic of a fermi-liquid, achieving a limiting value of $\chi_o \approx 63 \times 10^{-3}$ emu/mole-Ce near 2 K before ordering magnetically at lower temperatures.

In Fig. 2 we show the specific heat divided by temperature C/T of $Ce_3Pt_4In_{13}$ below 10 K and compare it to its non-magnetic analog $La_3Pt_4In_{13}$. The sharp peak at 0.95 K is indicative of a magnetic transition. Entropy considerations (discussed below) indicate that this feature stems from antiferromagnetic (AF) order in a doublet ground state. The up-turn in C/T below 200 mK is well described by a $C \sim T^{-2}$ term and as such we associate this with a nuclear Schottky contribution arising from the splitting of nuclear isospin levels in the presence of the internal field produced by the AF order. The data display a broad minimum at 5 K where C/T $\approx$ 630 mJ/mole-Ce $K^2$. In comparison, C/T for $La_3Pt_4In_{13}$ is smaller than the Ce data for all T < 20 K. A small anomaly is evident near 3 K, and this region is highlighted in a plot of C/T vs. $T^2$ shown in Fig. 3. The C/T anomaly centered a 3.3K coincides with a zero resistivity transition at the same temperature. In an applied field of 2 kOe this resistive transition drops to 2.5K, while 10 kOe is sufficient to move the transition below 1.5 K. From the 10 kOe data (Fig. 3) the Sommerfeld coefficient and Debye temperature for $La_3Pt_4In_{13}$ are $\gamma$ = 15 mJ/mole-La $K^2$ and $\theta_D$ = 186 K, respectively. The magnitude of the jump in C/T and $\gamma$ are related by $\Delta C/T_c = 1.95\gamma$. This value for $\Delta C/T_c$, coupled with the modest reduction in the resistive transition temperature in 2 kOe, signifies that this is an intrinsic superconducting transition and not a result of In flux contamination ($H_c$ = 293 Oe for In).[14] Hence, $La_3Pt_4In_{13}$ is a moderately strong-coupled 3.3 K superconductor. This $T_c$ is similar to those exhibited by the $(RE)_3M_4Sn_{13}$ stannides first reported by Remeika.[10]

The low-temperature magnetic state of $Ce_3Pt_4In_{13}$ can be ascertained by carefully examining the magnetic contribution to the specific heat below 20 K. $C_{mag}(T)$ is plotted in the inset to Fig. 2; $C_{mag}(T)$ is defined as $C_{mag} = C(T) - C_{lattice} - \alpha T^{-2}$, where $C_{lattice}$ is the lattice specific heat contribution of $La_3Pt_4In_{13}$ and the $T^{-2}$ term accounts for the nuclear Schottky upturn below 200 mK. There is a considerable magnetic contribution to $C(T)$ throughout the temperature region below 20 K, and the magnetic entropy approaches $2Rln2$ at 20 K. The cuboctahedral point symmetry at the rare-earth site splits the $J = 5/2$ Hund's rule multiplet of $Ce^{3+}$ into three doublets. The magnetic entropy below 20 K indicates that there are two doublets influencing $C(T)$ below 20 K with the third doublet well outside of this temperature window. The prominent broad peak in $C_{mag}(T)$ centered at 10 K is nicely described by a Schottky model involving two doublets separated by $\delta_1 = 25$ K; this Schottky contribution is denoted in the inset to Fig. 2 by the dashed line.

Removal of the Schottky contribution to $C_{mag}(T)$ clarifies the nature of the magnetic ground-state. The ground-state magnetic specific heat $[C_{mag} - C_{Schottky}]/T$ vs. T is presented in Fig. 4a, and the inset depicts the entropy $S(T)$ associated with the data. While the full doublet entropy $Rln2$ is present by 10 K, only 25% of $Rln2$ is released below $T_N$. This suggests that the ordered moment is heavily compensated by Kondo interactions. The remaining $(3/4)Rln2$ entropy is removed between $T_N$ and 10 K as reflected by the unusual extended tail in C/T above 1 K. The electronic contribution to the specific heat is difficult to determine because C/T does not saturate above $T_N$; a simple entropy balancing construction gives an estimated electronic contribution $\gamma \geq 1$ J/mole-C $K^2$. The Kondo energy scale $T_K$ for this $\gamma$ can be determined from the relation appropriate for a doublet ground state, $\gamma T_K = (\pi/6)R$.[15] This expression gives $T_K \approx 4.5$ K. This value is consistent with the slow, continuous rise in C/T that is evident between 5K and $T_N$. Below $T_N$, $C_{mag}(T)$ (see Fig. 4b) is best described by a model composed of a linear electronic contribution $\gamma_o T$ plus a term that accounts for the contribution from antiferromagnetic spin waves with the dispersion relation $w = \sqrt{\Delta^2 + Dk^2}$ ,[16]

$$C_{mag} = g_o T + a\Delta^{7/2}T^{1/2}e^{-\Delta/T}\left[1+(39/20)(T/\Delta)+(51/32)(T/\Delta)^2\right]. \tag{1}$$

In this expression $\Delta$ is the spin-wave gap and $\alpha$ is related to the spin-wave stiffness D by $\alpha \propto D^{-3/2}$. As shown in Fig. 4b, fitting Eqn. 1 to the $[C_{mag} - C_{Schottky}]$ data below $T_N$ yields an excellent fit with $\Delta = 1.05$ K, $\alpha = 1.9$ J/mole-Ce $K^4$, and a residual electronic contribution $\gamma_o = 360$ mJ/mole-Ce $K^2$.[17]. While a large remnant $\gamma$ is not uncommon in AF U-based heavy-fermions,[18] the same is not true for Ce systems.[19] We note that large $\gamma_o$ values in the magnetically ordered state have also been reported for two AF compounds that are isostructural to $Ce_3Pt_4In_{13}$; $Ce_3Ir_4Sn_{13}$ orders at $T_N = 2.0$ K and displays a residual Sommerfeld coefficient $\gamma_o = 670$ mJ/mole-Ce $K^2$[20], while $U_3Rh_4Sn_{13}$ orders at $T_N = 17.5$ K with a residual Sommerfeld coefficient $\gamma_o = 225$ mJ/mole-U $K^2$.[21]

Electronic transport measurements support the magnetic and Kondo configuration implied by the specific heat data. The temperature-dependent resistivity $\rho(T)$ of both $Ce_3Pt_4In_{13}$ (solid line) and $La_3Pt_4In_{13}$ (dashed line) from 100 mK to 325 K are presented in Fig. 5a. $La_3Pt_4In_{13}$ exhibits a resistivity that monotonically increases with increasing temperature while the Ce compound has a room-temperature resistivity of 150 $\mu\Omega$-cm and a positive d$\rho$/dT below 300 K. There is a shoulder in $\rho_{Ce}$ that is centered at 20 K and a precipitous drop is evident below 10 K. No Fermi-liquid behavior ($\rho \sim T^2$) is evident in $\rho_{Ce}$ above $T_N$. The inset shows the Ce compound's resistivity below 4 K; the AF transition manifests itself as an inflection point in $\rho$ at $T_N$. The resistivity due to electrons scattering with AF spin-waves described by the dispersion relation employed in fitting the specific heat is given by [16]

$$\rho(T) = \rho_o + \rho_{AF}\Delta^{3/2}T^{1/2}e^{-\Delta/T}\left[1+(2/3)(T/\Delta)+(2/15)(T/\Delta)^2\right], \quad (2)$$

where $\rho_o$ is a T-independent constant and $\rho_{AF} \propto D^{-3/2}$. Eqn. 2 fits the $T < T_N$ data extremely well with a spin-wave gap value ($\Delta = 1.1$ K) consistent with that obtained from the fit to the C(T) data. As for $T > T_N$, there is no evidence for Fermi-liquid behavior in the ordered state. The $Ce_3Pt_4In_{13}$ magnetic resistivity, $\rho_{mag} = \rho_{Ce} - \rho_{La}$, (the dotted line in Fig. 5) exhibits a temperature-dependence that is characteristic of a heavy-fermion compound: a negative $d\rho/dT$ above 50 K that varies logarithmically with temperature (see Fig. 6a), a broad maximum at intermediate temperatures, and a sharp drop at low temperatures. The shoulder in $\rho_{Ce}$ becomes a broad hump centered at $T_{max} = 12$ K in $\rho_{mag}$. The hump centered at $T_{max}$ is characteristic of the resistive anomaly produced in Kondo-lattice systems by a CEF level. A CEF-derived resistivity feature typically occurs near $T = \delta$,[22] and as such this feature is consistent with the Schottky anomaly splitting observed in C(T). It appears that the "coherence peak" commonly observed in the resistivity of Kondo lattice compounds coincides with the CEF resistivity feature.

The thermoelectric power S of $Ce_3Pt_4In_{13}$ for 5 K < T < 300 K is depicted in Fig. 5b. The TEP is positive at all temperatures, varies linearly with temperature above 100 K, and displays a peak with a maximum at 15 K. The TEP of Ce Kondo systems is composed of three terms, [23] (1) a linear-in-temperature term that stems from conventional carrier diffusion, (2) a positive term stemming from incoherent scattering from the CEF levels that peaks at roughly $T = \delta/3$, [24] and (3) a negative coherence peak that becomes large below $T_K$. The first two terms are clearly present in the S(T) data, and the peak at 15 K is qualitatively consistent with the 25 K

CEF splitting determined from the specific heat data. Although the data are always positive, the sharp drop below 15 K may be associated with the Kondo resonance term.

The temperature-dependent Hall coefficient (T > 5 K) measured in 10 kOe is also presented in Fig. 5b. As with the TEP, $R_H$ is positive and gradually increases with decreasing temperature below 300 K, where $R_H(300\ K) = 1.1 \times 10^{-10}$ m$^3$/C. No coherence-derived maximum is evident in the data above 5 K. The Hall coefficient in correlated Ce compounds is commonly described via a two-component model; the first term is T-independent and arises from conventional carrier-concentration effects: $R_o \sim 1/ne$, where n is the carrier concentration and e is the electron charge. The second term results from asymmetric skew scattering due to the large magnetic moment present on the Ce ions. Combined, the full Hall coefficient becomes [25]

$$R_H = R_o + \xi g \frac{\mu_B}{k_B} \frac{\chi(T)}{C} \rho_{mag}(T), \tag{3}$$

where $R_o$ is the conventional Hall term, g = 6/7 for full-moment Ce, $\mu_B$ is the Bohr magneton, $k_B$ is Boltzmann's constant, $\chi(T)$ is the T-dependent magnetic susceptibility, and C is the susceptibility's Curie constant. $\xi$ is a scattering constant that characterizes the strength of skew scattering; values of $\xi$ greater than 0.1 are physically unrealistic.[25] The solid line in Fig. 5b is a fit to Eqn. 3 with $R_o = 1.1 \times 10^{-10}$ m$^3$/C (corresponding to n ~ 6 × 10$^{22}$ holes/cm$^3$) and $\xi = 0.03$. The fit is comparable in quality to that obtained for many canonical heavy-Fermion systems.[25]

The influence of pressure P on the resistivity of Ce$_3$Pt$_4$In$_{13}$ is presented in Fig. 6. $\rho(T,P)$ was measured from 0.35K to 300 K in hydrostatic pressures ranging from 0.1 to 12.6 kbar, and the $\rho_{mag}(T,P)$ data are plotted with respect to a logarithmic T scale in Fig. 6a. At a pressure of 0.1 kbar, $\rho(T)$ exhibits the CEF-derived peak between 10 and 20 K, a $\rho \propto -ln(T)$ region between 20 K and 70 K, and a broad feature between 100 K and 300 K that may herald the presence of the third doublet CEF level located roughly 200 K above the magnetic ground-state ($\delta_2 \approx 200$ K). Increasing the pressure above 0.1 kbar monotonically increases the resistivity throughout the paramagnetic regime, and it also steadily increases the ln$T$ slope between 20 K and 70 K. The data in Fig. 6a indicate that pressure does not have a significant effect upon the temperature $T_{max}$ where the resistivity peaks near 20 K ($dT_{max}/dP = 12 \pm 22$ mK/kbar), nor on the location of the broad anomaly centered near 200 K. Since CEF-derived resistivity features are relatively insensitive to small volume changes, the pressure-insensitivity $\rho$ features at 20 K and 200 K are likely dominated by scattering from excited CEF levels. The influence of pressure on the ordering transition is the focus of Fig. 6b. Eqn. 2 continues to provide a good fit to the data below $T_N$ at all pressures to 12.6 kbar. Pressure acts to reduce the residual resistivity $\rho_o$ and increase the change in resistivity that occurs when the material is warmed above $T_N(P)$. The inflection point (maximum in $d\rho/dT$) that marks $T_N$ increases with pressure at the rate $dT_N/dP = 23$ mK/kbar.

The $\rho(T,P)$ data in Fig. 6a provide a means of deducing both the pressure dependence of the magnetic exchange energy, and degree to which the $T_N$ variation with pressure stems from $J_{ex}(P)$. The $\rho \propto -\ln(T)$ relationship for temperatures between 20 K and 70 K (Fig. 6a) occurs because the lower-lying doublet at $\delta_1 = 25$ K is almost fully occupied while the higher-lying doublet at $\delta_2 \approx 200$ K is only beginning to be populated in this temperature regime. Under these conditions, the slope $\Gamma \equiv d\rho/\ln T$ is given by [22]

$$\Gamma = \frac{4\bm{p}m^*v_o}{e^2\hbar} N(E_F)^2 J_{ex}^3 \frac{(l_i^2 - 1)}{(2J+1)} \quad , \tag{4}$$

where m* is the effective carrier mass, $v_o$ is the volume per Ce moment, J = 5/2 for Ce, and $\lambda_i$ is the degeneracy of the occupied CEF levels. Pressure-dependent values of $J_{ex}$ can be calculated by combining Eqn. 4 with experimental values for $\Gamma(P)$, and assuming that m*, $v_o$, and N($E_F$) = 2.1 eV$^{-1}$ [26] are pressure-independent. $J_{ex}$ varies linearly with P at the rate $dJ_{ex}/dP$ = 0.84 meV/kbar, and the estimated exchange energies at 0.1 and 12.6 kbar are -88.1 meV and -99.0 meV, respectively. Given the expression $T_N \propto J_{ex}^2$, the $J_{ex}(P)$ values can be used to derive the variation of the Néel temperature with pressure, and this can be compared to $T_N(P)$ as determined by tracking the onset of magnetic order in the resistivity. The estimated variation of $T_N(P) = T_N(0)J_{ex}^2(P)/J_{ex}^2(0)$ (dashed line in the inset to Fig. 6b) matches the $T_N$ values determined from the inflection point in $\rho(T,P)$ surprisingly well. The variation of $T_N$ with $J_{ex}$ for $Ce_3Pt_4In_{13}$ is $dT_N/dJ_{ex} \approx$ 27 mK/meV. This value for $Ce_3Pt_4In_{13}$ can be compared to $CeAl_2$ ($|dT_N/dJ_{ex}| \approx$ 130 mK/meV, $dJ_{ex}/dP$ = 0.33 meV/kbar) which orders antiferromagnetically near 4 K.[27,28] More generally, Ce Kondo systems exhibit $|dT_N/dJ_{ex}|$ and $dJ_{ex}/dP$ values ranging from 0 to 200 mK/meV, and 0.2 to 1.0 meV/kbar, respectively.[29] As such, $Ce_3Pt_4In_{13}$ exhibits a pressure-dependent magnetic exchange energy that is similar to other Kondo systems.

## IV.    Discussion

The ambient–pressure electronic transport, magnetic susceptibility, and specific heat data strongly support the viewpoint that $Ce_3Pt_4In_{13}$ is a heavy-fermion antiferromagnet. The limited entropy present at $T_N$ and the extended tail in C/T evident between $T_N$ and 10K are consistent with a $T_K$ of roughly 5 K. Despite these consistencies, the possibility remains that the tail results from critical fluctuations, disorder, or magnetic frustration. Critical fluctuations generally influence the specific heat only over a very narrow temperature range above $T_N$ [7] so it seems doubtful that fluctuations could produce the tail present in C/T. With regard to disorder, evidence for site exchange in the $Pr_3Rh_4Sn_{13}$ structure was recently observed in $Ce_3Ru_4Ge_{13}$,[30] but the key ingredient for this interpretation (a significantly reduced magnetic moment due to differing Ce site valences) is not present in $Ce_3Pt_4In_{13}$. Similarly, while frustration can led to an extended tail in C/T [31], the cuboctahedral Ce site symmetry in $Ce_3Pt_4In_{13}$ does not appear to be amicable to magnetic frustration. In the end, the experimental features that support a heavy fermion interpretation are (a) the entropy argument, (b) the large residual $\gamma_o$ present below $T_N$, (c) a fermi-liquid-like saturation in $\chi$ near 2 K corresponding to a Wilson ratio between $\gamma$ and $\chi_o$ of R $\approx$ 4, and (d) the precipitous drop in $\rho(T)$ below 20 K. While the importance of spin-wave scattering in the ordered state and the crystal-field anomaly centered at $T_{max}$ in $\rho(T)$ complicate a straightforward interpretation of the resistivity data (and presumably account for the absence of a $T^2$ Fermi-liquid regime in $\rho(T)$), the single-impurity Kondo-like resistivity above 50 K and the large drop in $\rho$ between $T_N$ and 20 K are consistent with strongly renormalized electronic state and an approach to coherence below $T_{max}$.

The positive $dT_N/dP$ evident in the $\rho(P,T)$ data is the one piece of experimental data that is difficult to reconcile with the heavy-fermion interpretation. The Doniach Kondo necklace model [3] generally provides a framework for qualitatively understanding the pressure response of Ce-compound.[32,33] In this model, a system with both magnetic order and a significantly enhanced $\gamma$ should display a negative $dT_N/dP$. That is, pressure should increase $|J_{ex}N(E_F)|$ in a heavy Ce compound, in turn increasing Kondo compensation of the local moments and lowering $T_N$. This is just what is observed in the $CeM_2Si_2$ (M = Ag, Au, Pd, and Rh) series where volumetric considerations lead to local-moment order at one end and heavy-fermion behavior at the other extreme.[33] The Doniach model also predicts a maximum in $T_N(P)$ which has been experimentally confirmed in moderately heavy electron magnets.[5] In contrast, a positive $dT_N/dP$ for $Ce_3Pt_4In_{13}$ combined with substantial thermodynamic and transport evidence for heavy electron behavior suggests that this material does not fit into the Doniach phase diagram in an obvious way. A determination of the effects of pressure on $T_K$ is required to confirm this suggestion. The presence of the crystal-field anomaly in $\rho(T)$ makes it impossible to unambiguously determine the influence of pressure on $T_K$ from $\rho(P,T)$ data. Hence, pressure-dependent specific heat measurements would be helpful in determining the effects of pressure on $T_K$ through the influence of pressure on $\gamma_o$ as would extending resistivity measurements to much higher pressures. Neutron quasi-elastic linewidth measurements, which are a more direct probe of $T_K$, combined with neutron diffraction measurements of the ordered moment would be helpful in clarifying our understanding of $Ce_3Pt_4In_{13}$ and its relationship to the Doniach model.

In summary, thermodynamic and ambient-pressure transport and magnetic measurements show that $Ce_3Pt_4In_{13}$ is a new heavy-fermion antiferromagnet with $T_N \approx 1$ K and $T_K \approx 5$ K. What makes this material a source of continuing interest is that a large residual Sommerfeld coefficient exits well below $T_N$, and that contrary to the Doniach phase diagram pressure acts to increase $T_N$. Neutron scattering and pressure-dependent thermodynamic measurements are warranted to further explore the ramifications of our observations. Clearly, the novel $Pr_3Rh_4Sn_{13}$ structure discovered more than 20 years ago by Remeika still deserves our attention.

**Acknowledgements**
   The authors thank P. Pagliuso for useful discussions. Work at Los Alamos was performed under the auspices of the U.S. Department of Energy. Z.F. acknowledge partial support from the N.S.F. (DMR-9870034 and DMR-9971348).

Figure Captions

Fig. 1. Inverse magnetic susceptibility plotted as a function of temperature for $Ce_3Pt_4In_{13}$. The solid line is a Curie-Weiss fit for T > 200 K, and the inset shows that $\chi$ saturates below 5 K.

Fig. 2. C/T vs. T for $Ce_3Pt_4In_{13}$ and $La_3Pt_4In_{13}$; the data are normalized per mole Ce or La. The inset shows $C_{mag}$ (J/mole-Ce K) vs. T for $Ce_3Pt_4In_{13}$; the dashed line is the specific heat due to a Schottky anomaly involving two doublets separated by 25K.

Fig. 3. The superconducting properties of $La_3Pt_4In_{13}$ are highlighted in this plot of C/T vs. $T^2$ in zero applied field (open triangles) and in 10 kOe (solid circles). The solid line is a Debye-model fit to the normal-state properties below 3.5 K.

Fig. 4. (a) Ground-state magnetic contribution to C/T plotted vs. T for $Ce_3Pt_4In_{13}$. The inset shows the magnetic entropy S plotted vs. T below 5 K with an arrow indicating $T_N$. (b) Ground-state magnetic C/T plotted vs. T focusing on the data in the ordered state. The solid line is a fit to the data using Eqn. 1 with $\Delta = 1.05$ K, $\alpha = 1.9$ J/mole-Ce $K^4$, and $\gamma_o = 360$ mJ/mole-Ce $K^2$.

Fig. 5. (a) Resistivity vs. temperature for $Ce_3Pt_4In_{13}$ (solid line), the non-magnetic analog $La_3Pt_4In_{13}$ (dashed line), and $\rho_{mag}$ for $Ce_3Pt_4In_{13}$ (the dotted line). The inset shows the resistivity in the vicinity of the magnetic ordering transition ($T_N$ is denoted by the arrow). (b) Temperature-dependent Hall effect $R_H$ (solid squares) and thermopower S (open circles) of $Ce_3Pt_4In_{13}$. The solid line is a fit to the Hall data that employs a skew scattering term.

Fig. 6. Pressure-dependent $\rho(T)$ for $Ce_3Pt_4In_{13}$. The pressures displayed include P = 0.1, 1.4, 5.6, 8.2, and 12.6 kbar. (a) $\rho_{mag}(T,P)$ data for T < 100 K; (b) data in the vicinity of $T_N$. The inset shows the pressure-dependent ordering temperature as determined by the location of the inflection point in $\rho(T)$. The dashed line represents an estimate of $T_N(P)$ determined from the pressure dependence of $J_{ex}$.

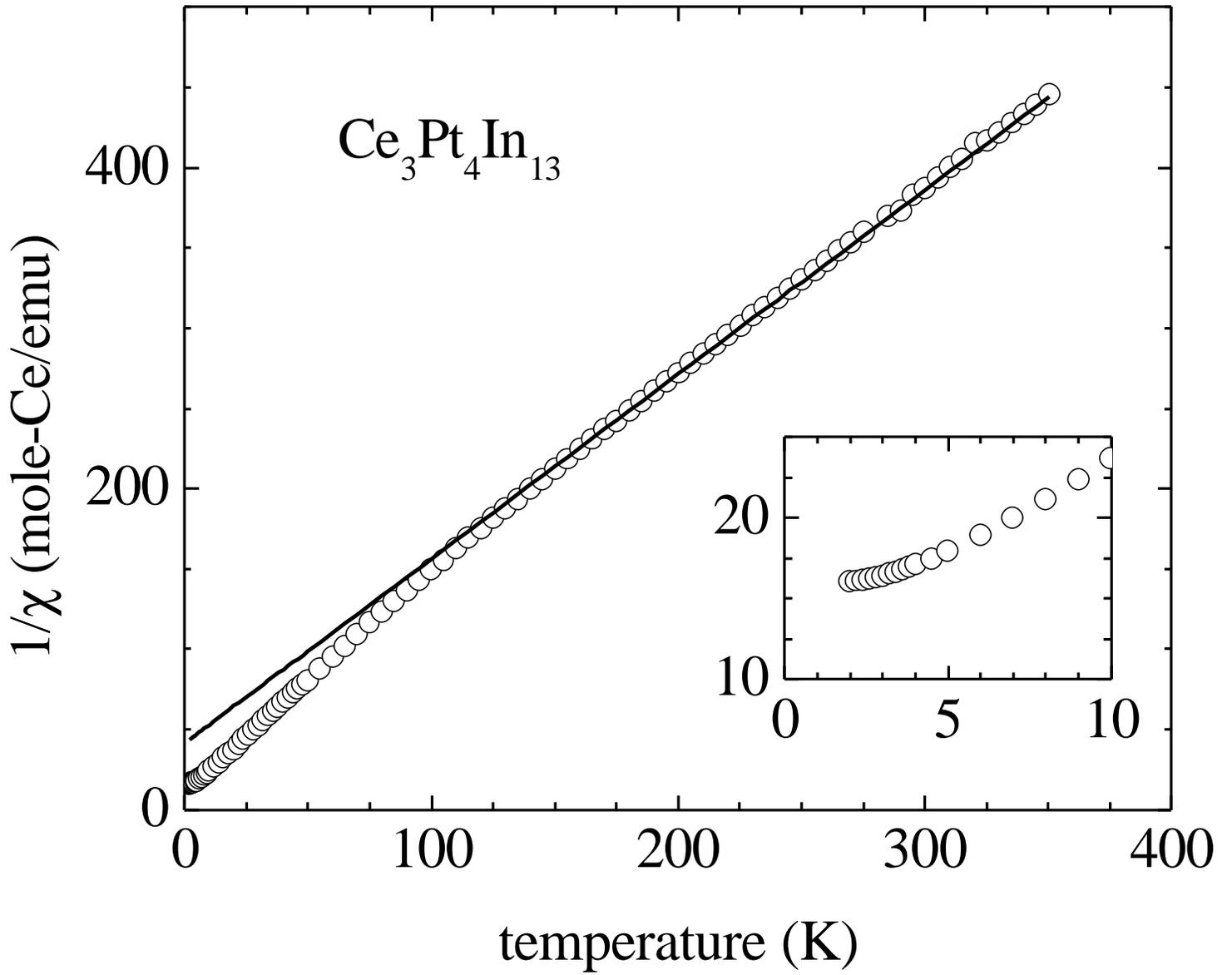

Hundley, *et al.* Figure 1

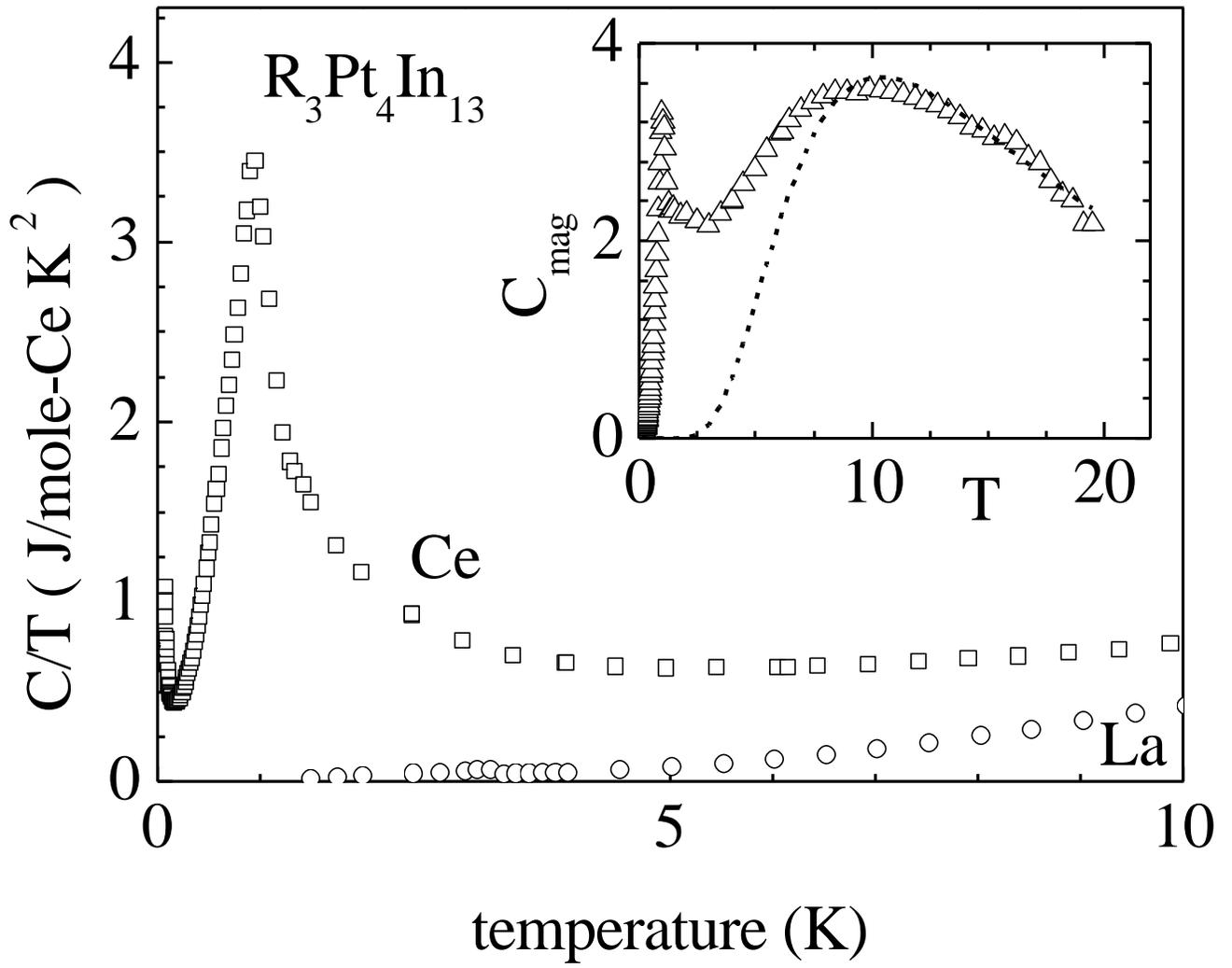

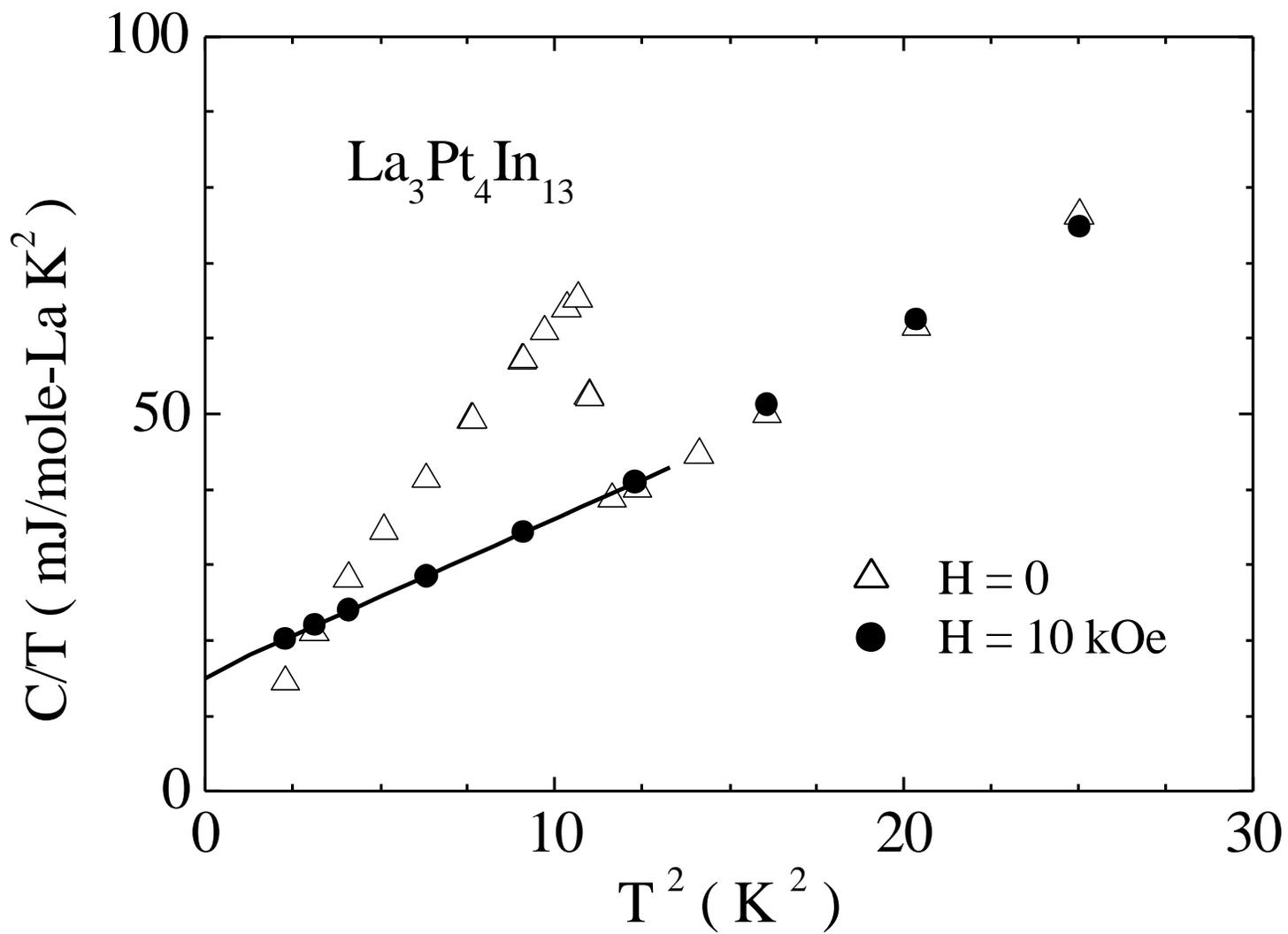

Hundley, *et al.* Figure 3

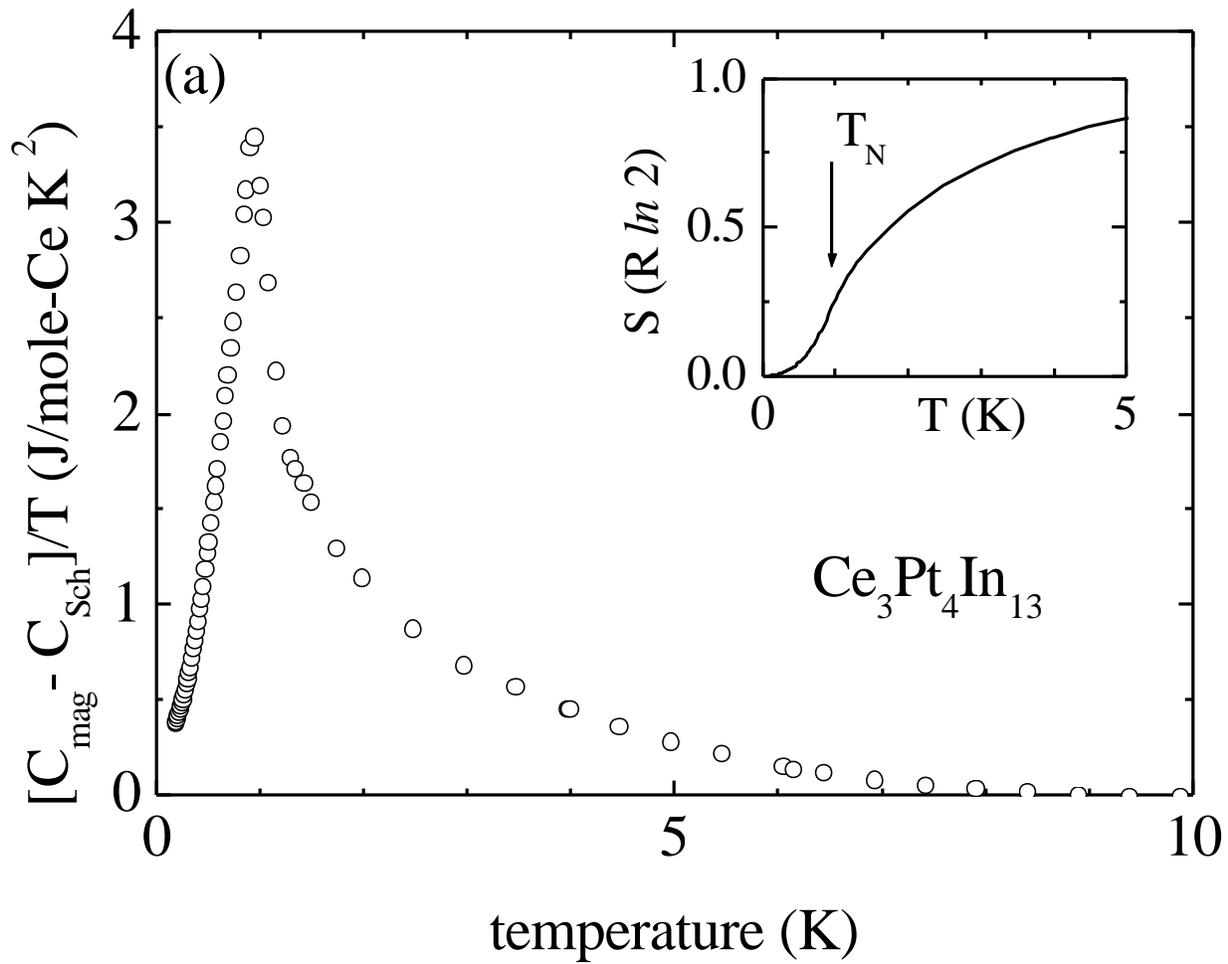

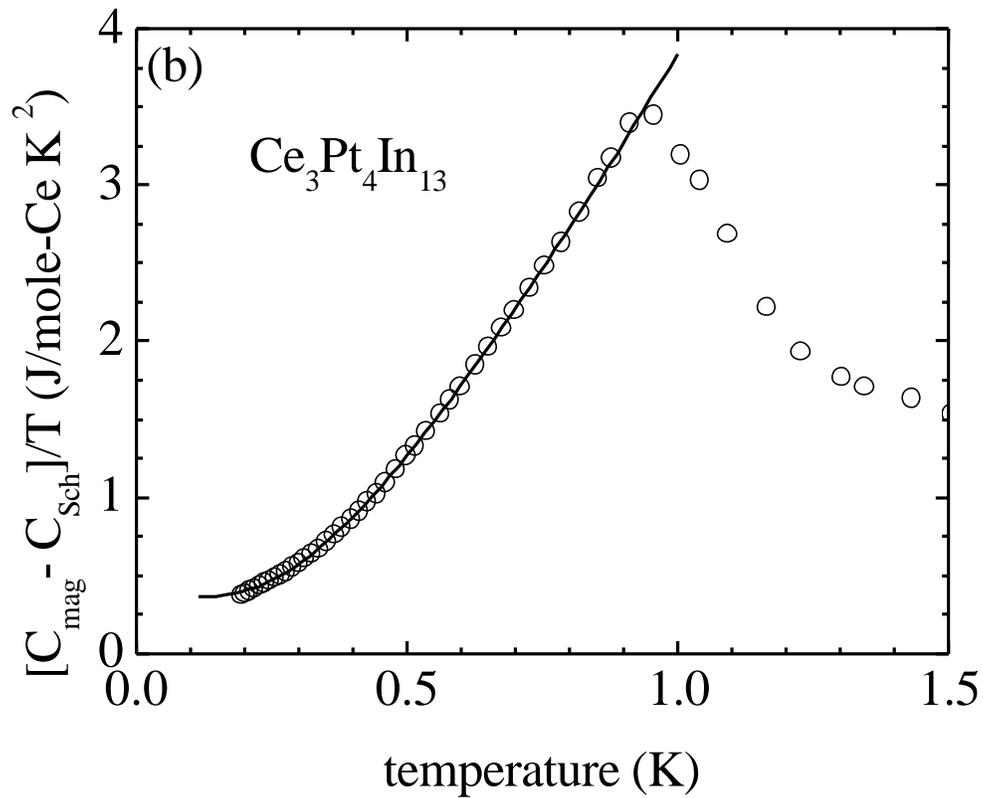

Hundley, et al. Figure 4

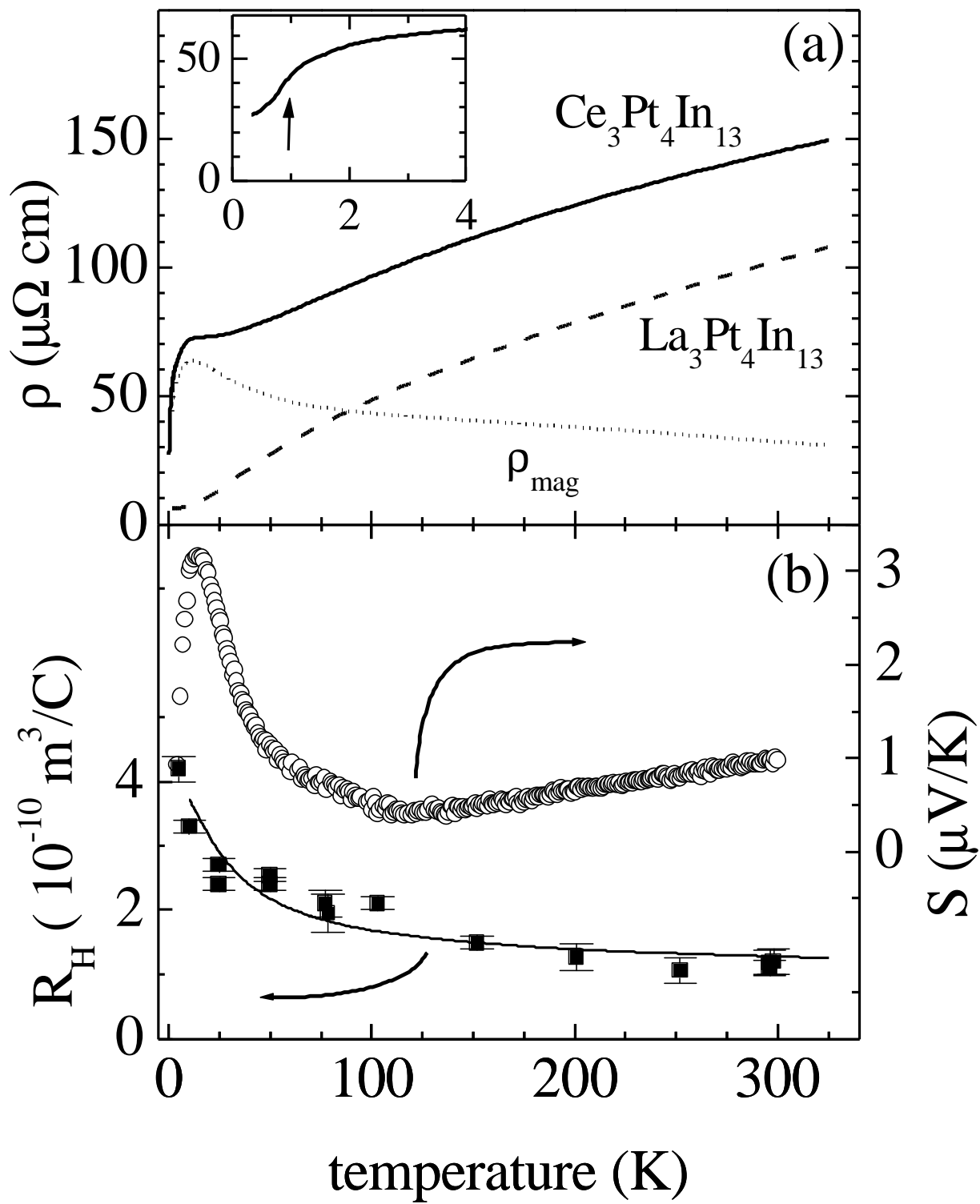

Hundley, *et al.* Figure 5

- 24 -

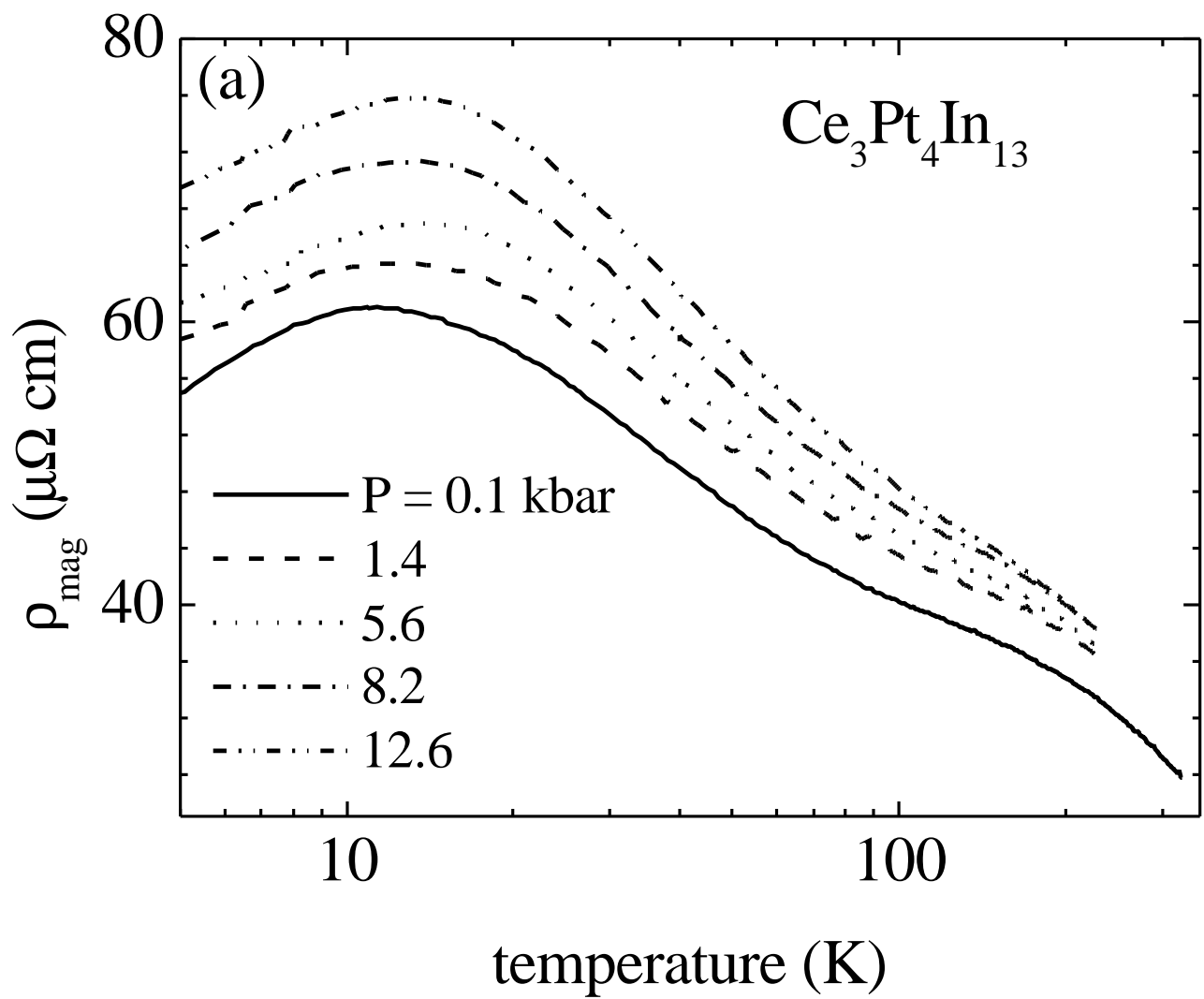

Hundley, *et al.* Figure 6a

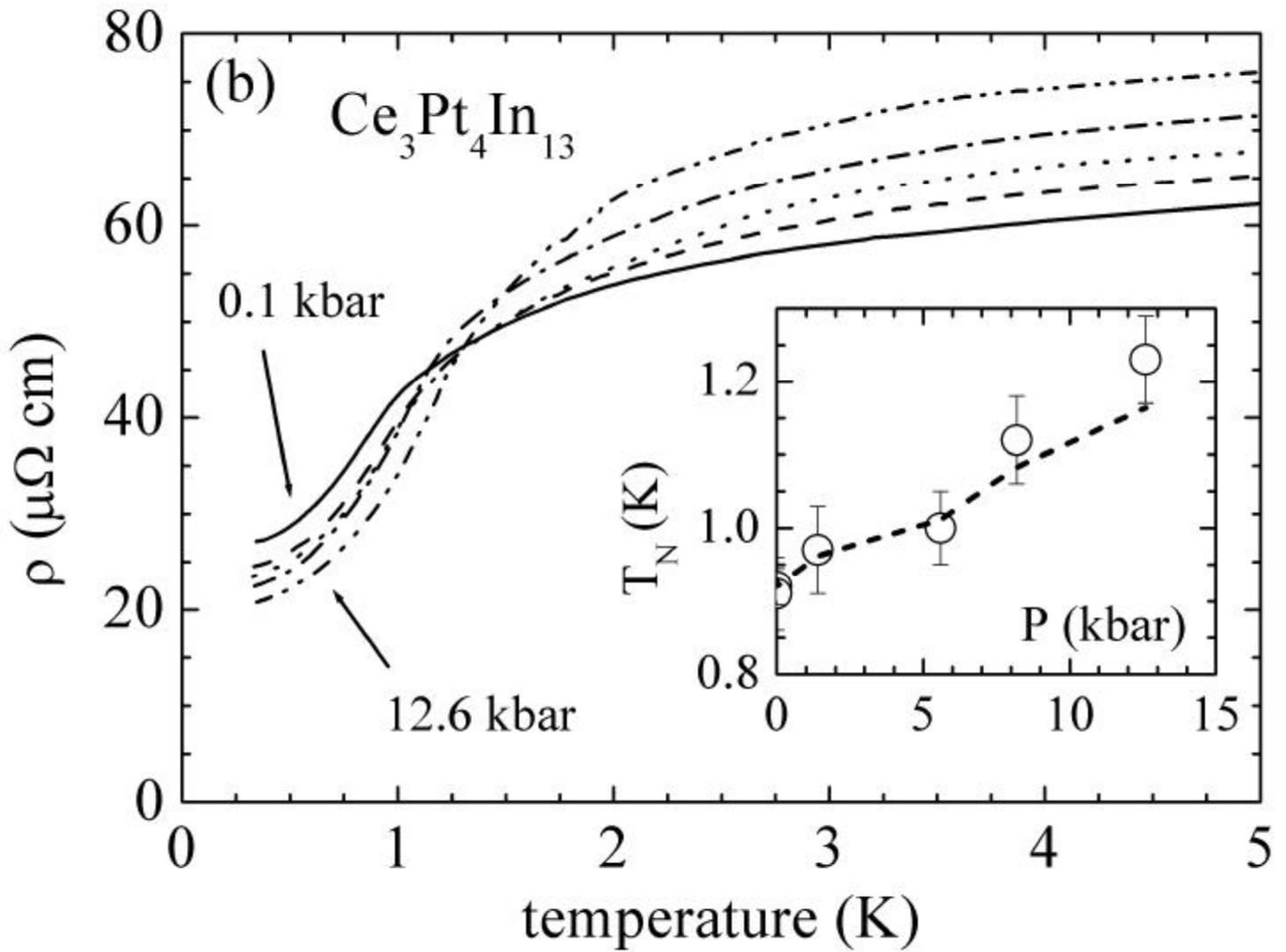

Hundley, *et al.* Figure 6b